\documentclass[twocolumn,superscriptaddress]{revtex4-1}

\usepackage{amsmath}
\usepackage{amssymb}
\usepackage{graphicx}
\usepackage{braket}
\usepackage{natbib}

\renewcommand{\r}{\mathbf{r}}
\newcommand{\F}{\mathbf{F}}
\newcommand{\R}{\mathbf{R}}
\newcommand{\n}{\mathbf{n}}
\renewcommand{\l}{\mathbf{l}}
\renewcommand{\S}{\mathbf{S}}
\newcommand{\shat}{\hat{s}}

\newcommand{\bomega}{\boldsymbol{\omega}}
\newcommand{\rhat}{\hat{r}}
\newcommand{\Q}{\mathbf{Q}}
\renewcommand{\t}{\mathbf{t}}
\newcommand{\bphi}{\boldsymbol{\phi}}
\newcommand{\C}{\mathbf{C}}
\renewcommand{\u}{\mathbf{u}}
\newcommand{\N}{\mathbf{N}}

\newcommand{\q}{\mathbf{q}}

\newcommand{\bdphi}{\delta \boldsymbol{\phi}}

\renewcommand{\P}{\mathbf{P}}
\newcommand{\Nhat}{\hat{n}}
\newcommand{\bchi}{\boldsymbol{\chi}}
\newcommand{\Sperp}{\mathbf{S}^{\perp}}
\newcommand{\lperp}{\mathbf{l}^{\perp}}
\newcommand{\bsigma}{\boldsymbol{\sigma}}
\newcommand{\btau}{\boldsymbol{\tau}}
\newcommand{\bOmega}{\boldsymbol{\Omega}}
\newcommand{\B}{\mathbf{B}}

\begin{document}

\title{Hidden symmetries generate rigid folding mechanisms in periodic origami}

\author{James McInerney}
\affiliation{School of Physics, Georgia Institute of Technology, Atlanta, GA 30332}
\author{Bryan Gin-ge Chen} 
\affiliation{Department of Physics and Astronomy, University of Pennsylvania, Philadelphia, PA, 19104}
\author{Louis Theran}
\affiliation{School of Mathematics and Statistics, University of St Andrews, St Andrews, Scotland}
\author{Christian Santangelo}
\affiliation{Department of Physics, University of Massachusetts Amherst, Amherst, Massachusetts 01003}
\affiliation{Department of Physics, Syracuse University, Syracuse, NY, 13244}
\author{Zeb Rocklin}
\affiliation{School of Physics, Georgia Institute of Technology, Atlanta, GA 30332}

\begin{abstract}
We consider the zero-energy deformations of periodic origami sheets with generic crease patterns.
Using a mapping from the linear folding motions of such sheets to force-bearing modes in conjunction with the Maxwell-Calladine index theorem we derive a relation between the number of linear folding motions and the number of rigid body modes that depends only on the average coordination number of the origami's vertices.
This supports the recent result by Tachi which shows periodic  origami sheets with triangular faces exhibit two-dimensional spaces of rigidly foldable cylindrical configurations. 
We also find, through analytical calculation and numerical simulation, branching of this configuration space from the flat state due to geometric compatibility constraints that prohibit finite Gaussian curvature.
The same counting argument leads to pairing of spatially varying modes at opposite wavenumber in triangulated origami, preventing topological polarization but permitting a family of zero energy deformations in the bulk that may be used to reconfigure the origami sheet.
\end{abstract}

\maketitle

\section{Introduction}
Origami-inspired materials are thin sheets whose two-dimensional crease patterns control their three-dimensional mechanical response, now manufacturable at the macroscopic scale using shape-memory alloys~\cite{hawkes2010, tolley2014} and the microscopic scale using graphene bilayers~\cite{miskin2018} or polymer films~\cite{bassik2009, liu2012, na2015}.
Origami principles are used to engineer deployable solar cells~\cite{miura1985}, stent grafts~\cite{kuribayashi2006}, flexible electronics~\cite{tang2014, song2014}, impact mitigation devices~\cite{ma2011}, and tunable antennas~\cite{liu2015} as well as characterize patterns in biological systems~\cite{mahadevan2005}.
Yet determining whether a crease pattern can be rigidly folded into a particular shape is an NP-hard problem~\cite{akitaya2018} due to nonlinear geometric constraints~\cite{belcastro2002} that can lead to  disjoint~\cite{liu2018} or branched~\cite{waitukaitis2015, stern2017, chen2018, berry2019} configuration spaces with multiple energetic minima~\cite{silverberg2014, silverberg2015}.

Periodic origami sheets yield uniform mechanical properties such as negative Poisson ratios~\cite{wei2013, schenk2013, lv2014, nassar2017, pratapa2019} and high stiffness-to-weight ratios~\cite{filipov2015}, making them apt for the design of mechanical metamaterials.
However, the study of origami tessellations has typically focused on crease patterns with inherent symmetries, such as the parallelogram faces of the Miura-ori~\cite{wei2013, schenk2013}, which both simplify their analysis and generate rigid folding motions~\cite{evans2015, waitukaitis2016, dieleman2019} that would  cost energy in the absence of these symmetries~\cite{pinson2017}.
One might naively expect such symmetries are required as triangulations of all convex polyhedra are rigid~\cite{gluck1975}.
However, Tachi recently found origami sheets composed of repeating unit cells with triangular but otherwise generic faces rigidly fold between cylindrical configurations, indicating that crease topology (the number of edges and vertices) may play as important  a role as crease geometry (the angles between these edges) in determining origami kinematics~\cite{tachi2015}.

In the present work, we similarly consider generic triangulations, which inform  the general case in three vital ways.
First, the rigidly foldable configurations of \emph{any} origami sheet can be derived as a subset of its triangulation's configurations.
Second, the \emph{low-energy} deformations of origami sheets are often well-approximated by the \emph{rigid} configurations of their triangulations~\cite{schenk2011, filipov2017}.
Finally, the triangulations are at the ``Maxwell point'': they have an equal number of constraints and degrees of freedom~\cite{lubensky2015,mao2018}, which we emphasize by calling them \emph{Maxwell origami}.
Mechanical systems at the Maxwell point generically  possess large numbers of both zero energy modes and force-bearing modes~\cite{guest2003, borcea2010} which can be localized to the boundary via topological polarization~\cite{kane2014, lubensky2015, mao2018}, provide directional response in the bulk~\cite{rocklin2016, rocklin2017njp}, and tuned by reconfigurations of the network~\cite{rocklin2017nat}.
However, origami sheets possess a geometrical duality between these two classes of modes~\cite{gluck1975, crapo1982, tachi2012} that, as we show, both permits the rigid foldability~\cite{tachi2015} and modifies its topological class, prohibiting the topological polarization~\cite{chen2016} of Maxwell origami which limits the ability to engineer directional response.

The remainder of the paper is organized as follows.
First, we review the work of Tachi to show Maxwell origami generically approximates a cylindrical sheet with two degrees of freedom~\cite{tachi2015}.
Next, we construct an index theorem that pairs folding motions with continuous symmetries in Maxwell origami.
We then show the restriction to cylindrical configurations leads to distinct branches of nonlinearly foldable origami configurations that we confirm through numerical simulation.
Finally, we extend our index theorem to accommodate spatially varying modes to explain the observed lack of topological polarization in Maxwell origami~\cite{chen2016}, and report lines of bulk modes with real wavenumber.

\begin{figure}[ht]
	\includegraphics[width=\columnwidth]{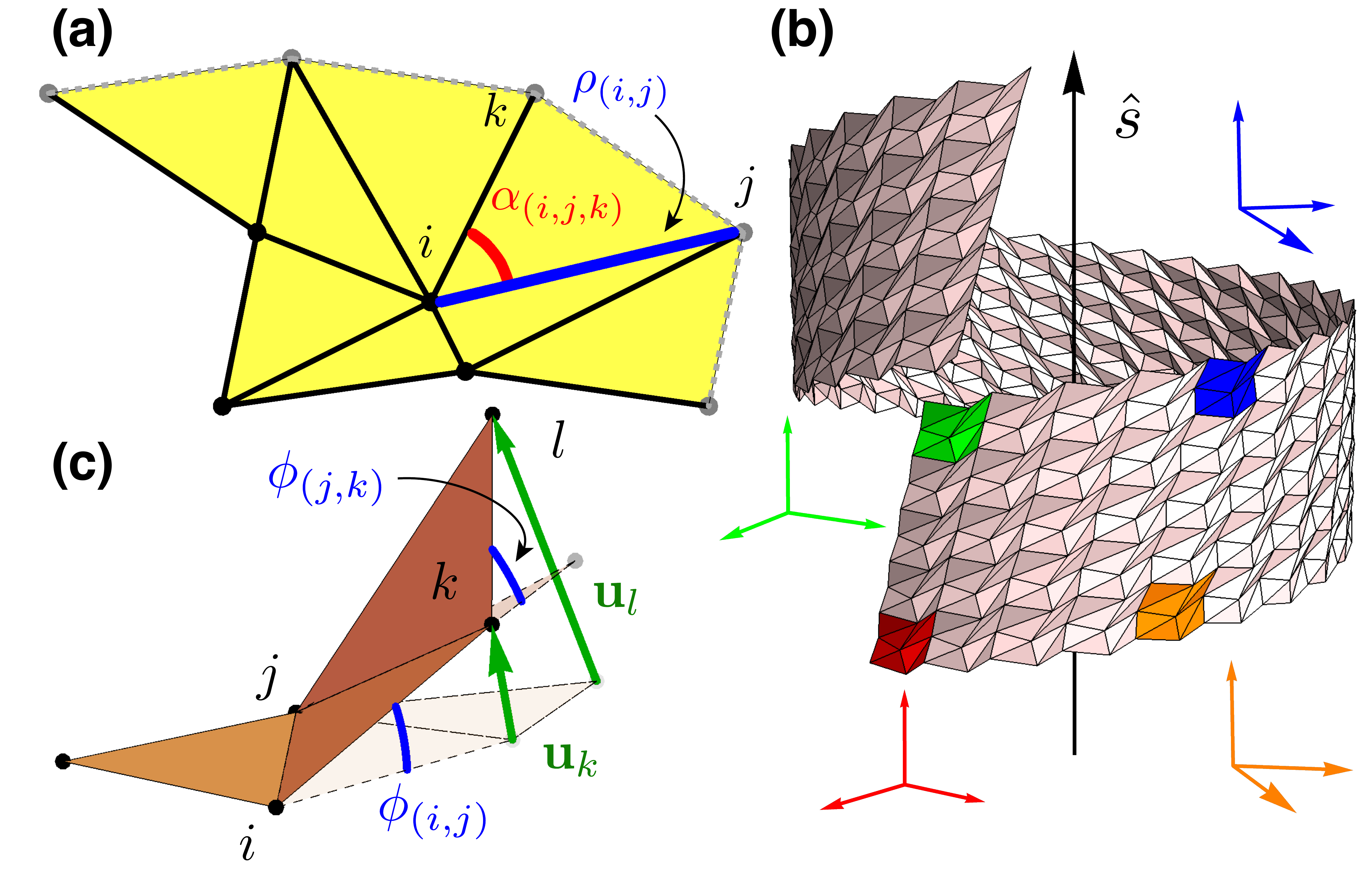}
	\caption{
(a) An origami unit cell with vertices labeled by Roman indices, sector angles labeled by $\alpha_{(i,j,k)}$, and fold angles labeled by $\rho_{(i,k)}$. 
(b) A generic, periodic origami sheet with cylindrical symmetries that follow from vertex compatibility with the unit cell highlighted in yellow. 
The screw periodicity of these sheets implies an orthonormal frame rotates between cells by the cell rotation matrices $\S_{1(2)}$.
(c) The zero modes of an origami sheet can be specified either by the vertex displacements, $\u_i$, on every vertex $i$ or the changes in folding angles, $\phi_{(i,j)}$ on every edge $(i,j)$.
The vertex displacements due to a folding motion accumulate allowing for non-zero curvature.
}
	\label{figsystem}
\end{figure}

\section{Cylindrical symmetries as a consequence of periodic origami angles}\label{secgeometry}
Origami sheets are parameterized by  fixed patterns of straight creases along which they can be folded rigidly, in the sense that no face bends or stretches.
This rigidity constraint determines what folded configurations are compatible with the underlying crease pattern.
Here, we introduce our notation and describe the most general origami structures with periodic folds, as previously explored by Tachi~\cite{tachi2015}.

We consider origami composed of unit cells, as depicted in Fig.~\ref{figsystem}a, with  sector angles $\alpha_{(i,j,k)}$ subtended by vertex-sharing edges $\r_{(i,j)}$ and $\r_{(i,k)}$, and fold angles $\rho_{(i,j)}$, given by the supplement of the respective dihedral angles, between pairs of adjacent faces (defined such that $\rho_{(i,j)}=0$ in a flat sheet), are identical in every cell, which are themselves indexed by $\n = (n_1, n_2)$. 
We do not require that the origami be developable, so that the sector angles need not sum to $2\pi$ around a vertex.

A necessary and sufficient condition for a set of fold angles to comprise a valid rigid fold about a vertex is for the successive rotations induced by traveling about the vertex to yield the identity rotation. 
This leads to the vertex condition derived in Appendix A~\cite{belcastro2002}

\begin{equation}\label{eqnbhconstraint}
\F_i = \prod_{(i,j)} \R_z(\alpha_{(i,j,k)}) \R_x(\rho_{(i,j)}) = \mathbf{I},
\end{equation}

\noindent where $j, k$ takes on the successive indices of vertices connected to vertex $i$ in counter-clockwise order and $\R_x, \R_z$ are matrices representing rotations about the $x$- and $z$-axes. For a simply connected sheet (as opposed to kirigami sheets with holes), this condition imposed at each vertex is sufficient to ensure rigid folding of the entire sheet.

Furthermore, the periodicity of sector angles between cells ensures that periodic fold angles can satisfy this condition in every cell. 
However, such periodic angles do not ensure that adjacent cells will have the identical orientations of normal crystalline structures. 
Instead, lattice rotation matrices

\begin{align}\label{eq:rotmat}
\S_{1(2)} \equiv \prod_{1(2)} \R(\rho_{(i,j)}, \rhat_{(i,j)}),
\end{align}

\noindent will relate the orientations between two faces in adjacent cells, where the products are taken over edges on paths between the two faces. 
This also means that the lattice vectors obtained by summing along edges, $\l_{1(2)} \equiv \sum_{1(2)} \r_{(i,j)}$ can only be defined in the first cell, and undergo rotations given by the lattice rotation matrices in other cells. 
Hence, in contrast with a conventional crystal whose cells are translations of one another along lattice vectors, the origami sheet is \emph{screw-periodic}: cells are related by screw motions consisting of translations and rotations (see Fig.~\ref{figsystem}b).

Any valid configuration, satisfying Eqn.~\ref{eqnbhconstraint}, must define unique relative orientations and positions of cells regardless of the path between them. 
Considering a loop between cells, such as the four colored cells in Fig.~\ref{figsystem}b, leads to the inter-cell position and compatibility conditions, 

\begin{align}
\S_1 \S_2 & = \S_2 \S_1, \label{eqnoricompat}\\
\l_1 + \S_1 \l_2 & = \l_2 + \S_2 \l_1. \label{eqnposcompat}
\end{align}
These conditions imply there is a unique rotation axis (except for flat sheets, and a few pathological cases that we will not consider), denoted by $\shat$, and a unique radius of curvature so that the sheet generically approximates a cylinder as shown in Fig~\ref{figsystem}b (see Appendix B for a characterization of this cylinder)~\cite{tachi2015}.
The familiar case of spatially periodic origami then emerges as the special limit in which the lattice rotations, $\S_{1,2}$, become identity matrices while arbitrary configurations with non-zero Gaussian curvature cannot be rigidly folded from periodic angles.
Given the position of each vertex in the origin cell, denoted by $\r_i$, we can compute the position of an arbitrary vertex by summation of all edge vectors traveled along to reach it
\begin{equation}\label{eqnvertexpos}
\r_i(\n) = \sum_{n'=0}^{n_1-1} \S_1^{n'} \l_1 + \S_1^{n_2} \sum_{n'=0}^{n_1-1} \S_2^{n'} \l_2 + \S_1^{n_1} \S_2^{n_2} \r_i,
\end{equation}
where the order of summation can be interchanged by orientation and position compatibility, Eqns.~\ref{eqnoricompat} and~\ref{eqnposcompat} (see Appendix C for an evaluation of the summations over lattice rotations).

Those compatibility conditions allow a prediction for the dimension of the space of cylindrical configurations of a triangulated origami. 
Consider a potential configuration specified by the positions of each vertex, the two lattice vectors, and the two lattice rotation matrices. 
A rigidly folded configuration of the triangulation must preserve the length of each edge and satisfy position and orientation compatibility. 
Euler's polyhedron formula states that the numbers of faces, edges and vertices must satisfy $N_v - N_e + N_f = \chi$, where the Euler characteristic $\chi$ vanishes for a doubly periodic surface.
Every face in a triangulation has three edges, each shared with exactly one face so that $N_e = (3/2)N_f$, thereby implying $N_e = 3 N_v$. 
In this way, each three-dimensional vertex position is accounted for via three edge constraints. 
Additionally, there are twelve numbers that specify the lattice vectors and lattice rotation matrices. 
The compatibility conditions supply four constraints: that the direction of the axis of the second rotation is shared by that of the first, and that two components of the position vectors in Eq.~\ref{eqnposcompat} are equal (the third direction, along the shared axis, is guaranteed to be equal). 
This leaves an eight-dimensional space of configurations of the sheet, six of which are simply rigid rotations and translations, leaving a \emph{two-dimensional} space of rigidly foldable deformations. 
This was observed by Tachi~\cite{tachi2015}, who advanced a similar counting argument. 
We will see these deformations emerge explicitly by considering higher-order rigidity conditions, which also reveal subtle branching behavior around the flat state.
Helical, cylindrical tubes have the two additional constraints that the ratios of lattice rotation angles, $2 \pi \theta_1 / \theta_2$, and on-axis components of the lattice vectors $l_1^{\shat} / l_2^{\shat}$ are rational to ensure closure, which generically renders the tubes rigid.
Allowing the tube to close without vertices connecting relieves the second of these conditions and permits motion by slip with a single degree of freedom~\cite{feng2020}.

\section{Linear folding motions from global symmetries via vertex duality}\label{seclinfolding}

\subsection{Relationship between folding angles and vertex displacements} 
The cylindrical symmetries of origami correspond to rigid body modes which are paired with force-bearing modes at the Maxwell point~\cite{maxwell1864, calladine1978}.
These force-bearing modes, however, are identical to infinitesimal changes in the fold angles, $\phi_{ij}$, which satisfy the vertex condition, Eqn.~\ref{eqnbhconstraint}, to first-order~\cite{gluck1975, crapo1982, tachi2012}.
Here, we combine this mechanical duality with the mechanical criticality of Maxwell origami to show rigid body modes generate linear folding motions independent of the sector angles.

Consider infinitesimal changes (zero modes) $\phi_{(i,j)}$ to the fold angles $\rho_{(i,j)}$. The linearization of Eqn.~\ref{eqnbhconstraint}, as shown in~\cite{tachi2009} and recapitulated in Appendix D, is

\begin{equation}\label{eqnlinfold}
\sum_{(i,j)} \phi_{(i,j)}(\n) \hat{r}_{(i,j)}(\n) = \mathbf{0},
\end{equation}
where the edges rotate $\hat{r}_{(i,j)}(\n) = \S_1^{n_1} \S_2^{n_2} \hat{r}_{(i,j)}$ between cells by Eqn.~\ref{eqnvertexpos}. 
The infinitesimal rotation of a face $(i,j,k)$ may be described by an ``angular velocity'' vector $\bomega_{(i,j,k)}$ such that any vector $\mathbf{v}$ on the face, including edge vectors, undergoes a rotation $\mathbf{v} \rightarrow \mathbf{v} + \bomega_{(i,j,k)} \times \mathbf{v}$ as shown in Fig.~\ref{figsystem}c. 
Two faces sharing an edge must then induce the same rotation upon it, leading to a relation between adjacent angular velocities and the folding angle of the edge between them:

\begin{equation}\label{omegadiff}
\bomega_{(i,j,l)}-\bomega_{(i,j,k)}= \phi_{(i,j)}\hat{r}_{(i,j)}
\end{equation}

\noindent These then accumulate such that the angular velocity of one face relative to a fixed face is
\begin{equation}\label{eqnangvel}
\bomega_{(i,j,k)}(\n) = \sum_{(i',j')} \phi_{(i',j')}(\n') \hat{r}_{(i',j')}(\n'),
\end{equation}
where the sum is over all edges crossed on a path between the faces.
Similarly, the displacement of a vertex on a distant face is given by the sum of all vertex displacements along the path from a fixed vertex, which are in turn determined by rotating the bond vectors via their respective angular velocities:
\begin{equation}\label{eqnvertexdisp}
\u_k(\n) = \sum_{(i',j')} \bomega_{(i',j',k')}(\n') \times \r_{(i',j')}(\n').
\end{equation}
This summation is explicitly evaluated for both the spatially periodic and screw periodic cases in Appendix G.

Having described how vertex positions may be generated via arbitrary folding motions, we may complete the identification by  a map from the vertex positions of an isometry back to the folding motions. The procedure is to take two edge vectors along a face, $\r_{(i,j)}, \r_{(k,j)}$ and the normal vector $\N_{(i,j,k)} = \r_{(i,j)} \times \r_{(k,j)}$ and to consider the changes implied by the vertex displacements to the two vectors $\u_{(i,j)}, \u_{(k,j)}$ and to the normal vector $\delta \N_{(i,j,k)} = \r_{(i,j)} \times \u_{(k,j)} + \u_{(i,j)} \times \r_{(k,j)}$ This yields the matrix equation

\begin{equation}\label{getomega}
\bomega^{\times}_{(i,j,k)} \left( \r_{(i,j)} \hspace{5pt} \r_{(k,j)} \hspace{5pt}  \N_{(i,j,k)}\right) = \left( \u_{(i,j)} \hspace{5pt}  \u_{(k,j)} \hspace{5pt}  \delta \N_{(i,j,k)}\right)
\end{equation}

\noindent which may be inverted to obtain $\bomega^{\times}_{(i,j,k)}$, the cross-product matrix whose elements give the angular velocity of the face. 
From these angular velocities Eqn.~\ref{omegadiff} may be used to obtain the changes to the folding angles.

\subsection{Duality between folding motions and tensions}

The linear folding constraint, Eqn.~\ref{eqnlinfold}, takes the familiar form of tensions $t_{(i,j)}$ along edges $\hat{r}_{(i,j)}$ that yield no net force called \emph{states of self stress}~\cite{gluck1975,crapo1982,tachi2012}.
 This hidden symmetry between static and kinematic modes has particular significance for periodic sheets, for which it implies symmetrically distributed edge modes, as we discuss later. 
The concatenation of Eqn.~\ref{eqnlinfold} at each vertex within the origin cell yields the equilibrium matrix, $\Q$, that maps tensions to the net force on each vertex.
Importantly, the static-kinematic duality reveals that the transpose of the equilibrium matrix is the compatibility matrix, $\C = \Q^T$, that maps vertex displacements to bond extensions~\cite{calladine1978}. 
This leads, via the rank-nullity theorem of linear algebra, to the celebrated Maxwell-Calladine index theorem relating the number of zero energy vertex displacements, $N_{zm}$, to the number of states of self stress, $N_{ss}$, within the origin cell~\cite{maxwell1864,calladine1978}
\begin{equation}\label{eqnindexthm}
N_{zm} - N_{ss} = 3 N_v - N_e.
\end{equation}

We are now able to combine the criticality of triangulated origami, which ensures the right-hand side vanishes, with the duality between states of self stress and folding modes to use spatial symmetries to guarantee the existence of folding modes, some of which have already been observed. 
Spatially periodic sheets have three translational modes, implying three states of self stress and three folding motions, as observed in triangulations of the Miura-ori and eggbox crease patterns~\cite{wei2013, schenk2013, lv2014, nassar2017, pratapa2019}. 
In contrast, cylindrical sheets have only two rigid-body modes: translations along and rotations about the axis, implying the two linear motions that lead to the two-dimensional space of configurations. 
In either case, fusing two triangular faces together to create a quadrilateral face eliminates a degree of freedom, reducing the space of rigid configurations.
More generally, folding motions are possible in origami above the Maxwell point due to symmetries, e.g. the Miura-ori, that render constraints degenerate as has been observed in spring networks~\cite{guest2014}.

\section{Nonlinear constraints lead to branching between cylindrical configurations}\label{secnonlinear}

In this section, we describe the full set of nonlinear rigid folds of the origami sheets. 
Spatially periodic states have three linear modes and we employ second-order rigidity conditions to identify how they extend to the nonlinear branches.
As we show, the necessary requirement that the linear modes generate a cylindrical surface is sufficient for a second-order folding motion to exist. 
The surface of modes in configuration space (parametrized by the fold angles) is generally two-dimensional, with two two-dimensional branches connected at the flat state.
In contrast, in developable sheets up to $2^{N_v+1}$ branches can meet at the flattened state, with every sheet investigated showing pairs of branches distinguished by whether a vertex pops upward or downward, as previously observed in origami sheets with one-dimensional configuration spaces~\cite{chen2018}.

While first-order compatibility is sufficient to ensure a cylindrical configuration folds into another cylindrical configuration (see Appendix G), the lattice rotation axes spontaneously chosen when folding from a spatially periodic state are not necessarily coaxial.
We can see this by noting the expansion of orientation compatibility, Eqn.~\ref{eqnoricompat}, about the flat state, where the lattice rotations are identity matrices, is trivially satisfied to first-order.
Instead, the leading order contribution is given
\begin{equation}\label{eqnquadconstraint}
\delta \S_1 \delta \S_2 = \delta \S_2 \delta \S_1,
\end{equation}
where the $\delta \S_{1,2}$ are skew-symmetric generators of rotation whose components are given by the inter-cell angular velocity $\sum_{1,2} \phi_{(i,j)} \rhat_{(i,j)}$ computed from Eqn.~\ref{eqnangvel} (see Appendix G).
From position compatibility, Eqn.~\ref{eqnposcompat}, we have at first-order $\delta \S_1 \l_2 = \delta \S_2 \l_1$, implying these rotations lie in the plane of the origami sheet defined by $\l_1 \times \l_2$ so that Eqn.~\ref{eqnquadconstraint} has only a single nonzero entry.
Taking linear combinations of the linear folding motions, $\phi_{(i,j)} = \sum_{\alpha} \lambda_{\alpha} \phi_{(i,j)}^{\alpha}$, this becomes a quadratic expression in the real coefficients, $\lambda_{\alpha}$, which will generically admit two distinct families of solutions, $\lambda_{\alpha}^{\pm}$, that correspond to upwards or downwards folded cylinders.
We note real solutions to Eqn.~\ref{eqnquadconstraint} do not always exist, as is the case for the triangulated Miura-ori, which prevents its out-of-plane linear motions from extending nonlinearly~\cite{wei2013}, however we find real solutions generically exist for our Maxwell origami sheets without any fine-tuning.

That the linear folding motions yield a cylindrical configuration turns out to be a sufficient condition for the existence of second-order folding motions, $\delta \phi_{(i,j)}$, which satisfy the vertex constraint, Eqn.~\ref{eqnbhconstraint}, to second-order.
This second-order vertex condition consists of a linear term in $\delta \phi_{(i,j)} \rhat_{(i,j)}$ and a quadratic sum of pairwise products of $\phi_{(i,j)} \hat{r}_{(i,j)}$ over each edge connected to a particular vertex $i$ (see Appendix E for an expansion of Eqn.~\ref{eqnbhconstraint})
\begin{equation}\label{eqnquadfold}
\sum_{(i,j)} \delta \phi_{(i,j)} \hat{r}_{(i,j)} + \sum_{(i,j)} \phi_{(i,j)} \bigg( \sum_{\substack{(i,k) \\ k<j}} \phi_{(i,k)} \hat{r}_{(i,k)} \bigg) \times \hat{r}_{(i,j)} = \mathbf{0},
\end{equation}
where $k<j$ denotes the interior sum is taken over successive indices clockwise from $j$ up to the starting edge.
The interior sum of the second term gives, by Eqn.~\ref{eqnangvel}, the angular velocity of a face relative to the starting face at vertex $i$ so that the cross product gives the rotation of edge $\rhat_{(i,j)}$ with the first edge of the sum held fixed.
By the first-order condition, Eqn.~\ref{eqnlinfold}, we can add any constant angular velocity, $\bomega_i$, to this sum since the exterior sum $\bomega_i \sum \phi_{(i,j)} \times \rhat_{(i,j)}$ vanishes, allowing us to rewrite Eqn.~\ref{eqnquadfold} as
\begin{equation}\label{eqnsecondorder}
\sum_{(i,j)} \delta \phi_{(i,j)} \hat{r}_{(i,j)} + \sum_{(i,j)} \phi_{(i,j)} \delta \rhat_{(i,j)} = \mathbf{0},
\end{equation}
where $\delta \rhat_{(i,j)}$ depends on the coefficients, $\lambda_{\alpha}$, used to construct the linear folding motion and are themselves linear in the $\phi_{(i,j)}$.
This means when we concatenate Eqn.~\ref{eqnsecondorder} at each vertex, the first term is the action of the equilibrium matrix on the second-order folding motions, $\Q \bdphi$, while the second term is the action of the \textit{change} in the equilibrium matrix due to a linear folding motion on the linear folding motions, $\delta \Q \bphi$, where we use bold to denote the vector of fold angle changes $\bphi = ( \ldots, \phi_{(i,j)}, \ldots )$.

Since we have already restricted our linear folding motions to those which yield cylindrical configurations, the compatibility matrix of the linearly deformed state, $\C' = \C + \delta \C$, must admit zero modes, $\u'$, corresponding to translations and rotations about the uniquely defined axis.
These are paired with states of self stress, $\t'$, that lie in the nullspace of the new equilibrium matrix, $\Q' = \Q + \delta \Q$, via mechanical criticality which, by the mechanical duality, are isomorphic to linear folding motions $\bphi'$.
Such new linear folding motions can generically be written as a combination of the linear folding motions in the original configuration, $\bphi$, along with a correction, $\bdphi$, that satisfies Eqn.~\ref{eqnsecondorder} after dropping the higher-order term $\delta \Q \bdphi$. 
Hence, the existence of the second-order folding motions of Eqn.~\ref{eqnsecondorder} is guaranteed so long as the first-order motions generate a cylindrical surface. 
As shown explicitly in Appendix F, this result can also be derived via the mechanical duality, which reveals a connection between rigid translations and rotations.

Finally, let us consider developable origami sheets in the flat state which admit extra linear folding motions (the generalization to origami sheets with both developable and non-developable vertices is straightforward).
This can be seen by noting Eqn.~\ref{eqnlinfold} only furnishes two constraints per vertex when all edges lie in a plane.
These additional folding motions are paired with zero modes that correspond to vertices popping up or down out of the plane~\cite{chen2018,berry2019}.
Generally, this yields an extra $N_v - 1$ linear folding motions for developable origami in the flat state (the rigid-body translation in the direction normal to the sheet can be written as a linear combination of the $N_v$ additional modes arising from developability) which do not all extend to rigid folding motions.
The $N_v$ seemingly missing constraints are provided by the quadratic term in Eqn.~\ref{eqnquadfold}.
Since every edge lies in the same plane, this yields a single constraint per vertex.
Moreover, this term is in the left nullspace of the equilibrium matrix so no $\delta \phi_{(i,j)}$ are needed to satisfy Eqn.~\ref{eqnbhconstraint} to second-order.
We generalize our definition of sector angles so that $\alpha_{(i,j,k)}$ is the angle between edges $\r_{(i,j)}$ and $\r_{(i,k)}$ which do not necessarily share a face but are coplanar. 
Eqn. \ref{eqnquadfold} then simplifies to the scalar equation for a developable vertex $i$

\begin{equation}\label{eqndevquadfold} 
\sum_{(i,j)} \sum_{\substack{(i,k) \\ k>j}} \phi_{(i,j)} \phi_{(i,k)} \sin(\alpha_{(i,j,k)}) = 0. 
\end{equation}

\noindent By taking linear combinations of our folding motions, we can find simultaneous solutions to the $N_v$ second-order constraints.
While there are up to $2^{N_v+1}$ complex roots by B\'{e}zout's theorem~\cite{harris2013, chen2018}, we are only interested in the real-valued solutions whose existence depends on the crease geometry. 
Since a developable sheet has reflection symmetry through the plane of the sheet, these roots come in pairs which fold upwards or downwards into indistinguishable cylinders.
In other words, for $N$ branches there are only $N/2$ unique branches which cannot be obtained by rotations of the remaining $N/2$ branches.

\section{Numerical investigation of nonlinear folding}

\begin{figure*}[t]
    \includegraphics[width=2\columnwidth]{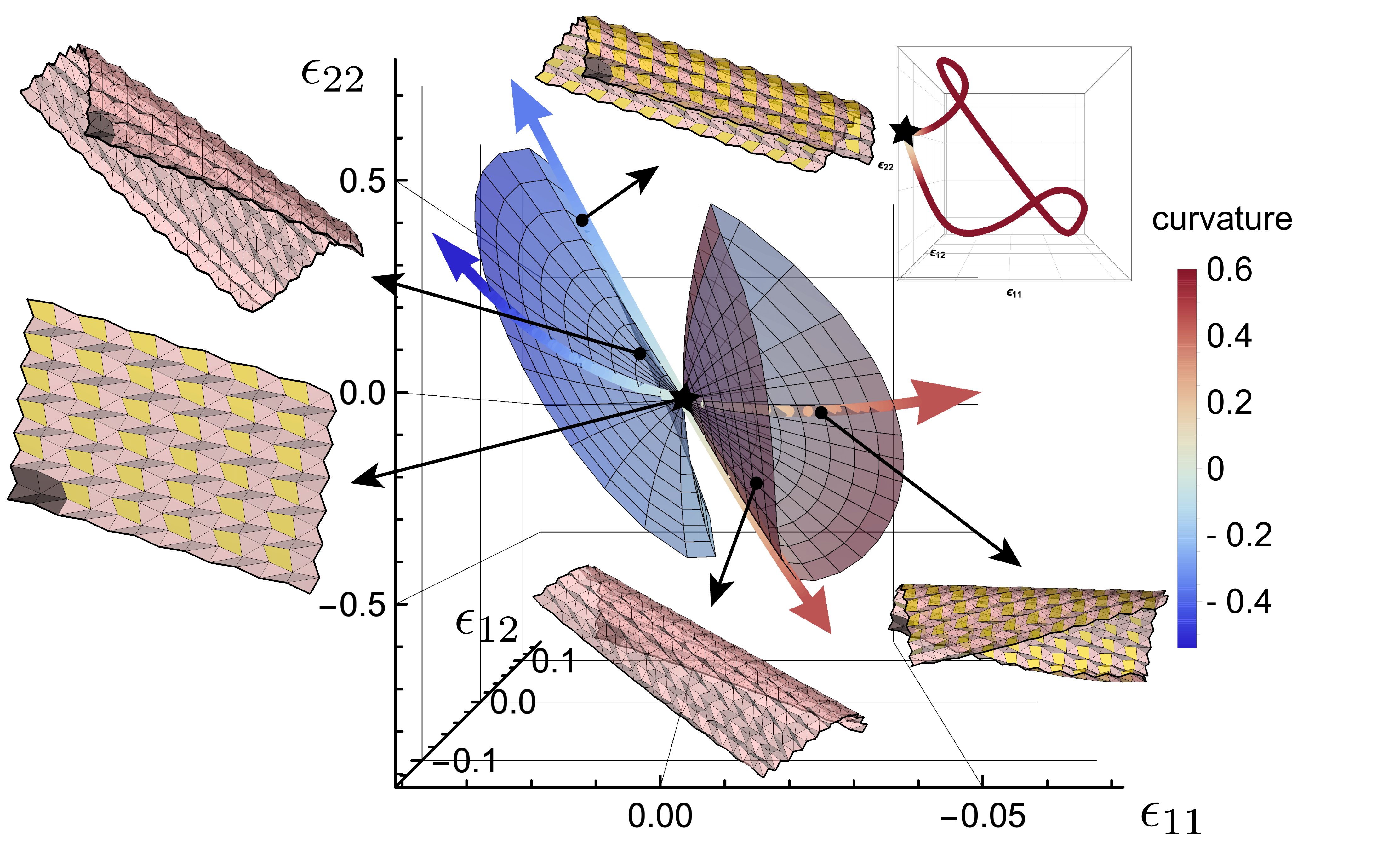}
    \caption{The 2-dimensional surface of rigidly foldable configurations for a nondevelopable triangulated origami sheet projected from its $N_e$-dimensional configuration space to the 3-dimensional strain space (see Appendices H and I) where coloring indicates the signed radius of curvature at each point.
Arrows point from a point in this space to the corresponding reconstructed sheet with cell $\n = (0,0)$ colored in gray. 
Some origami sheets have two triangular faces fused into a rigid quadrilateral marked in yellow, restricting the folding motions from the full 2D surfaces to one-dimensional paths marked with curved, multi-colored arrows.
The yellow quadrilateral indicates a polygonal face which restricts the sheet to 1-dimensional trajectories.
The distinct branches correspond to origami sheets which fold upwards or downwards from the flat state.
The boxed inset shows two one-dimensional folding trajectories as they close into a single loop over high strains.
}
    \label{fignondev}
\end{figure*}

We now show corrections exist at all orders by numerically evolving periodic origami sheets.
We begin with a spatially periodic, non-developable origami sheet composed of six triangular faces and a single quadrilateral face in each cell (this unit cell with four vertices is the simplest pattern with no trivial creases), as labeled by the star in Fig. \ref{fignondev}, and rigidly fold along its one-dimensional branches.
Following, we add a crease across the diagonal of the quadrilateral face, allowing the sheet to explore its two-dimensional space of rigidly foldable configurations embedded in the $N_e$-dimensional configuration space.
Alternatively we may obtain a one-dimensional path through the two-dimensional configuration space by locking the fold angle on any edge whether or not the adjacent faces form a polygon.
To visualize this surface, we project into a three-dimensional space spanned by strains of the lattice vectors.
We use the three independent components, $(\epsilon_{11}, \epsilon_{22}, \epsilon_{12})$, of the in-plane deformation tensor determined via changes to the lengths and angle between the lattice vectors as described in Appendix I.

\begin{figure*}[t]
    \includegraphics[width=2\columnwidth]{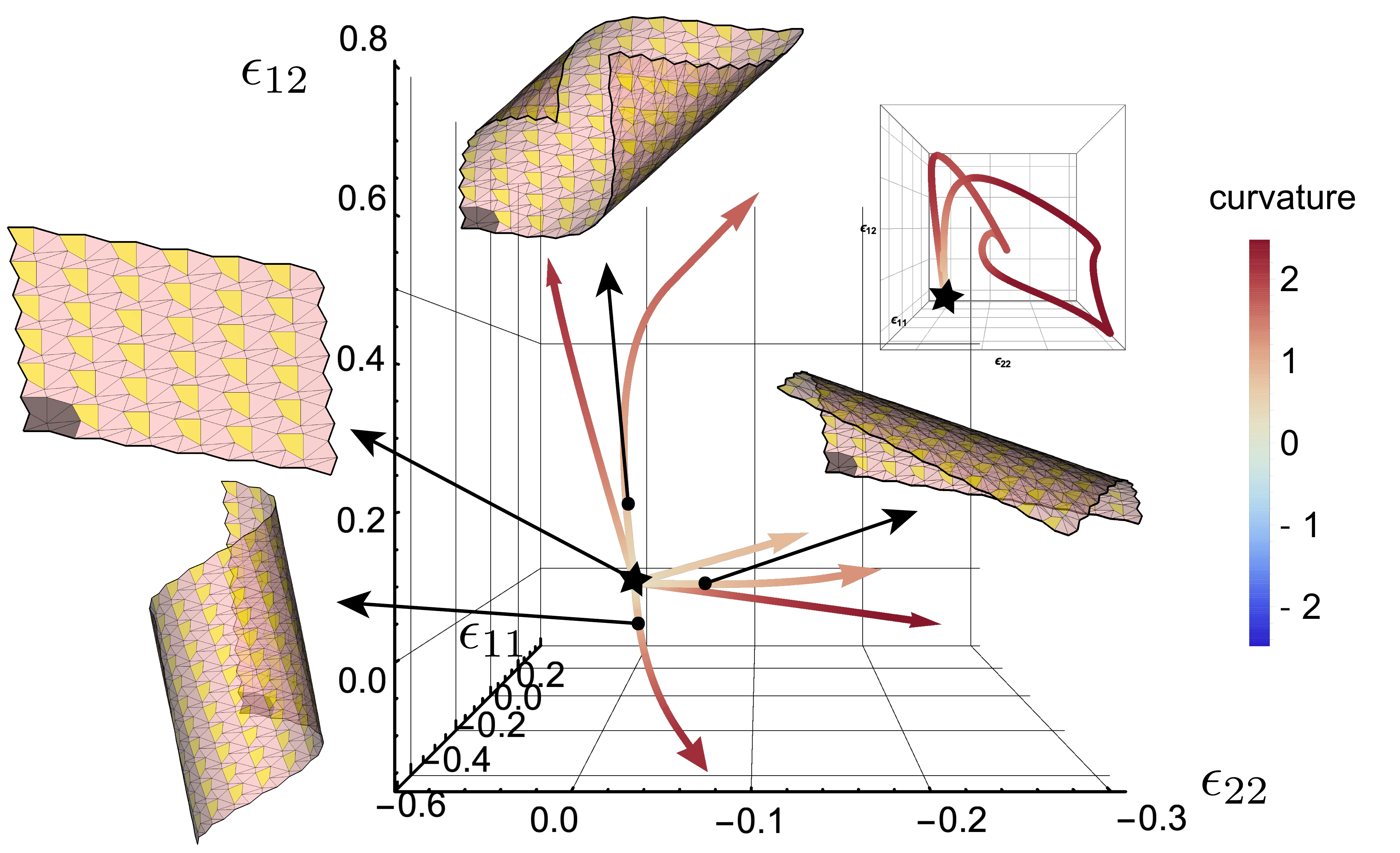}
    \caption{The 1-dimensional lines of rigidly foldable configurations for a developable origami sheet with one quadrilateral face per unit cell
projected from its $N_e$-dimensional configuration space to the 3-dimensional strain space (see Appendices H and I) where coloring indicates the signed radius of curvature at each point where the flattened configuration is labeled by a star.
Arrows point from a branch in this space to the corresponding reconstructed sheet with cell $\n = (0,0)$ colored in gray and the insets show a full one-dimensional orbit through the configuration space when constraining the fold angle on an edge.
Our randomly generated crease pattern admits $6$ solutions to Eqn. \ref{eqndevquadfold} and hence has $3$ branches with strictly positive radii of curvature.
There are $3$ additional branches with identical strains and the oppositely signed curvatures.
}
    \label{figdev}
\end{figure*}

In Fig. \ref{fignondev}, we show the branched one-dimensional paths and the two-dimensional surfaces corresponding to configurations of the origami sheet along with a spatial configuration of the sheet on each branch (see Supplementary Movie 1 for evolution along this surface).
The branches are colored according to the configuration's radius of curvature at each state, where the sign is chosen to designate whether the sheet folded upwards or downwards.
In Appendix H, we discuss how the spatial embedding and curvature direction of these configurations is obtained from the fold angles.
These trajectories close when allowing for self-intersection of the origami sheet (some fold angles  pass through $\pm \pi$ at which point adjacent faces intersect), as shown in the inset of Fig.~\ref{fignondev} which, although unphysical for origami, may have consequences in the behavior of equivalent systems such as spin origami~\cite{roychowdhury2018prl}.
Although the two-dimensional surfaces close, we only show a closed one-dimensional path as otherwise features are obscured by spurious self-intersections due to different configurations with the same in-plane strains despite having distinct fold angles.

We next construct a developable origami sheet with a single quadrilateral face in the flat state.
Our arbitrarily chosen crease pattern yields six real solutions to Eqn. \ref{eqndevquadfold}, indicating six branches from the flat state.
We show three of these branches in Fig.~\ref{figdev} (each branch has a beginning and end which join in the flat state), all with positive radius of curvature (see Supplementary Movie 2 for evolution along this surface).
The remaining three branches have the exact same in-plane strains with equal and opposite radii of curvature.
The number of branches is a property of the crease geometry and we do not address a method for controlling the number of branches here.
In fact, even identical triangulations with different faces fused into a quadrilateral substantially effects which strains and curvatures (geometry) occur in addition to the number of branches (topology).
For developable origami, the lattice vectors are maximal in the flattened state so any folding results in $\epsilon_{11}, \epsilon_{22} < 0$, while shearing allows for either positive or negative values of $\epsilon_{12}$.

\section{Pairing of spatially varying modes at opposite wavenumbers}

In the previous sections we considered the pairing of rigid-body modes and deformations with the same fold angle changes in every cell. 
Here, we generalize the mechanical duality to spatially varying modes to investigate the topological mechanics of Maxwell origami whose connections to quantum mechanical systems such as topological insulators~\cite{kane2014}, nodal semimetals~\cite{rocklin2016}, dissipative systems~\cite{gong2018topological} and spin origami~\cite{shender1993kagome,chandra1993anisotropic,ritchey1993spin,roychowdhury2018prl} are discussed in Appendix J.
These spatially varying zero modes are normal modes of the system (with frequency zero) and so, due to Bloch's theorem, must take the forms

\begin{equation}
\u_i(\n) = \u_i z_1^{n_1} z_2^{n_2}, \hspace{5pt}    
\phi_{(i,j)}(\n) = \phi_{(i,j)} z_1^{n_1} z_2^{n_2},
\end{equation}

\noindent for Bloch factors $z_i = e^{i q_i}$ with wavenumbers $q_i$ which may be complex for general boundary conditions. 
The mapping from vertex displacements to folding motions in Eqn.~\ref{eqnvertexdisp} extends naturally to the spatially varying modes, which inherit the same dependence on wavenumber so that, by the mechanical duality of origami, there is a mapping between zero modes and states of self stress at finite wavenumber $N_{zm}(\q) = N_{ss}(\q)$.

\begin{figure}
	\includegraphics[width=\columnwidth]{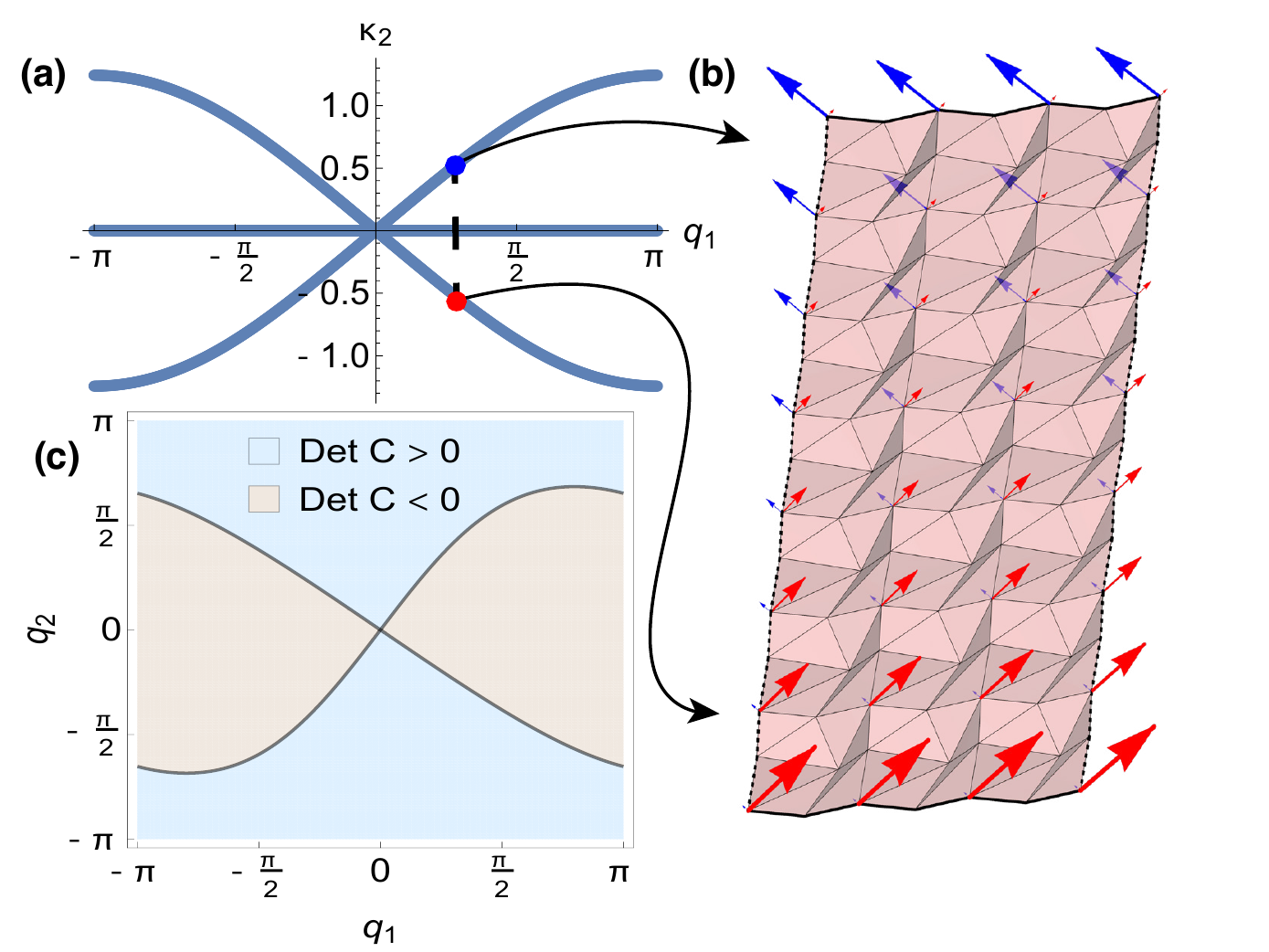}
	\caption{
(a) The signed inverse decay lengths, $\kappa_2$, along the $\l_2$ direction for zero modes at particular, real assignments of $q_1$.
The line along the origin corresponds to bulk zero modes that have zero inverse decay length everywhere.
(b) The spatially periodic origami sheet shown in (a) with blue (red) arrow indicating zero mode vertex displacements on one vertex per cell (additional arrows omitted for visual clarity) that grow towards the top (bottom).
(c) The determinant of the compatibility matrix for the origami sheet in panel (b) across the Brillouin zone.
}
	\label{fignonpolarization}
\end{figure}

The finite wavenumber static-kinematic duality relates the equilibrium matrix to the transpose of the compatibility matrix at the \emph{opposite} wavenumber, $\Q(\q) = \C^T(-\q)$, modifying the Maxwell-Calladine index theorem of Eqn.~\ref{eqnindexthm} to~\cite{lubensky2015, mao2018}
\begin{equation}\label{eqnfiniteindex}
N_{zm}(\q) - N_{ss}(-\q) = 3 N_v - N_e,
\end{equation}
which pairs zero modes at $\q$ with self stresses at $-\q$ (this sign difference, crucial for our argument, has been omitted previously). 
This leads to the intriguing scenario, identified by Kane and Lubensky~\cite{kane2014}, in which a zero mode may be exponentially localized to one edge (at some complex $\q$) with a state of self stress at the opposite edge (at $-\q$), creating an excess or deficit of zero modes on an edge or interface beyond that predicted by local coordination number, which is known as topological polarization.
The localization of these modes can be characterized by an inverse decay rate $\kappa_{1(2)} = - \text{Im}(q_{1(2)})$, where, e.g. $\kappa_2 < 0 (\kappa_2 > 0)$ indicates the mode is exponentially localized on the bottom (top) edge as shown in Fig~\ref{fignonpolarization}a-b.

Since these states of self stress can themselves be mapped onto zero modes via the duality discussed above for triangulated surfaces, whenever there is a zero mode at $\q$ there must also be one at $-\q$, as shown by the pairing of inverse decay rates in Fig.~\ref{fignonpolarization}a.
This means while it is always possible to impose a periodic distortion on a surface and, by the fundamental theorem of algebra, find a mode that exponentially decays into the bulk, the hidden symmetry guarantees that there is a corresponding mode on the opposing side.
This shows polarization can never occur in Maxwell origami as observed by Chen et al.~\cite{chen2016}.
In fact the same work found Maxwell kirigami, composed of equal numbers of quadrilateral faces and holes, to topologically polarize is reconciled by a generalized version of the mechanical duality which pairs folding motions of the original structure with the self stresses of a distinct structure obtained by replacing all faces with a hole and vice versa~\cite{finbow2012}, thereby breaking the hidden symmetry.

Interestingly, this characteristic of Maxwell origami, while eliminating the Kane Lubensky invariant, generates a new topological property. 
The determinant of the compatibility matrix becomes a Laurent polynomial in the Bloch factors, $\det \C(\q) = \sum_{m,n} c_{m n} z_1^m z_2^n$, where the highest-order of $m$ and $n$ is given by the total number of edges passing from the unit cell to the $\n = (1,0)$ and $\n = (0,1)$ cells respectively and $c_{m n}$ are real coefficients determined by the crease geometry. 
This determinant vanishes at wavenumbers admitting zero modes, and previously it has been shown in 2D Maxwell lattices~\cite{rocklin2016} that the real and imaginary parts of the compatibility matrix generically vanish at zero-dimensional points within the 2D Brillouin Zone. 
In the present case, the existence of zero modes at equal and opposite wavenumbers implies this determinant must be purely real so there instead appear 1D lines of zero modes, as shown in Fig.~\ref{fignonpolarization}a, corresponding to the lines of magnetic waves observed in a quantum analog of origami sheets~\cite{roychowdhury2018prl}. 
Furthermore, the sign of the real compatibility matrix serves as a topological invariant which changes only when crossing such a line of zero modes as shown in Fig.~\ref{fignonpolarization}c.

It is not clear how these linear modes extend into nonlinear deformations. 
By mechanical criticality, a triangulated sheet with open boundaries must have  modes due to the missing constraints at the edges. 
In general, though, the existence of such finite-wavenumber modes is guaranteed only by continuous symmetries that are broken as the mode is extended nonlinearly.
In particular finite-wavenumber modes will induce sinusoidally varying amounts of Gaussian curvature through the sheet in contrast to the uniform folding motions that extend nonlinearly.

\section{Conclusion}\label{secconclusion}
We have considered the rigid foldability of periodically triangulated origami with generic crease patterns and constructed a counting argument via Maxwell origami's combined mechanical duality and mechanical criticality. 
That argument shows translational and rotational rigid-body modes ensure the existence of folding motions that extend nonlinearly to yield two-dimensional spaces of rigidly foldable origami configurations which branch from the spatially periodic configuration.  
Furthermore, we showed this allows construction of crease patterns with a single degree of freedom simply by adding a single quadrilateral face to the unit cell.
We leave for future work the refinement of our counting argument to address how discrete symmetries can permit non-triangulated patterns, such as the Miura-ori, to rigidly fold.

Finally, we have extended our counting argument to spatially varying modes, revealing that edge modes necessarily appear in pairs on opposite sides, explaining the lack of polarization previously observed~\cite{chen2016}.
Our analysis reveals the existence of one-dimensional lines of bulk zero modes in Maxwell origami, as opposed to zero-dimensional points, that could be used to reconfigure the origami sheet by introducing an expanded unit cell. This also identifies a new topological invariant based on this hidden symmetry between folding motions and states of self stress that may lead to additional topological properties~\cite{roychowdhury2018prb}.
The generality of our results is unique in that it depends only on the coordination of the crease pattern rather than the specific geometry which may aid in the design of foldable materials in hard-to-control, microscopic environments.

B.G.C. thanks funding from NSF Award No. PHY-1554887. C.S. thanks funding from NSF DMR-1822638. J.M. thanks support from the Georgia Institute of Technology President's Fellowship and the STAMI Graduate Student Fellowship.

\bibliographystyle{unsrt}
\bibliography{HiddenSymmetriesBib.bib}

\appendix
\onecolumngrid

\section{Vertex and loop constraints}
Here we provide a proof of the Belcastro-Hull vertex constraint for generic vertices.
Let us condense our notation from the main text and label the $N_e$ edges leaving a vertex as $\rhat_i$ for $i$ taking integer values from $1$ to $N_e$.
We similarly denote the normal vectors on each face as $\Nhat_i$, where the $i$th face precedes the $i$th edge.
Then taking $\rhat_1 = \hat{x}$ and $\Nhat_1 = \hat{z}$, the rotation $\P_1 = \R_z(\alpha_1) \R_x(\rho_1)$ defines the similarity transform that rotates the coordinate basis to $\rhat_2 = \hat{x}$ and $\Nhat_2 = \hat{z}$ via $\P_1^{-1} \rhat_1 \P_1 = \rhat_2$ and $\P^{-1} \Nhat_1 \P_1 = \Nhat_2$.
Now suppose we are at the $i$th edge and have found a similarity transform such that $\rhat_i = \hat{x}$ and $\Nhat_i = \hat{z}$.
Then the rotation $\P_i = \R_z(\alpha_i) \R_x(\rho_i)$ defines the similarity transform that rotates the coordinate basis to $\rhat_{i+1} = \hat{x}$ and $\Nhat_{i+1} = \hat{z}$ via $\P_i^{-1} \rhat_i \P_i = \rhat_{i+1}$ and $\P_i^{-1} \Nhat_i \P_i = \Nhat_{i+1}$.
Hence for any edge we can define the similarity transform $\bchi_i = \prod_{i'=1}^i \P_{i'}$ to rotate the initial edge and face normals to the $i$th edge and face normals $\rhat_i = \bchi_i^{-1} \rhat_1 \bchi_i$ and $\Nhat_i = \bchi_i^{-1} \Nhat_1 \bchi_i$.
Since we have chosen $\rhat_1 = \hat{x}$ and $\Nhat_1 = \hat{z}$, we must have $\bchi_i = \mathbf{I}$ for a closed loop around the vertex yielding
\begin{equation}
\F_i = \prod_{(i,j)} \R_z(\alpha_{(i,j,k)}) \R_x(\rho_{(i,j)}) = \mathbf{I},
\end{equation}
after returning subscripts as in the main text.

\section{Cylindrical surfaces from periodic origami sheets}
Here we show how inter-cell orientation and position compatibility, 

\begin{align}
\S_1 \S_2 & = \S_2 \S_1, \label{eqnoricompat}\\
\l_1 + \S_1 \l_2 & = \l_2 + \S_2 \l_1, \label{eqnposcompat}
\end{align}

\noindent can be used to define a continuous cylinder.
First, we fix the two lattice rotation matrices, $\S_{1,2}$, to be rotations about the $\shat$ axis to enforce orientation compatibility, Eqn.~\ref{eqnoricompat}.
It immediately follows the projection of position compatibility, Eqn.~\ref{eqnposcompat}, along this rotation axis is trivially satisfied.
We then examine the components transverse to $\shat$ denoted by the superscript $\perp$
\begin{equation}\label{eqnperpposcompat}
\lperp_1 + \Sperp_1 \lperp_2 = \lperp_2 + \Sperp_2 \lperp_1.
\end{equation}
Here, $\Sperp_{1,2}$ denote two-dimensional rotation matrices which map transverse vectors to transverse vectors so Eqn.~\ref{eqnperpposcompat} is invertible, requiring the lattice vectors satisfy
\begin{equation}\label{eqnperpposcompatsol}
\lperp_2 = (\mathbf{1} - \Sperp_1)^{-1}(\mathbf{1} - \Sperp_2) \lperp_1.
\end{equation}
Since we can always choose a Cartesian basis which has one axis along $\shat$ and another along $\lperp_1$, Eqns.~\ref{eqnoricompat} and~\ref{eqnposcompat} imply our lattice is specified by the five-parameter family of $\{\theta_1, \theta_2, \lvert \lperp_1 \rvert, \l_1 \cdot \shat, \l_2 \cdot \shat\}$.

Now suppose we would like to define a cylinder which intersects the first vertex in every cell.
Clearly, this cylinder's symmetry axis must lie along the origami's axis of rotation.
To obtain its radius, we again consider the projection of the surface into the plane defined by this axis.
This surface connects the first vertex of the unit cell with its counterpart in the neighboring cells by a planar curve.
The transverse lattice vectors, $\lperp_{1,2}$, are then the geometric chords of this curve which subtend the corresponding rotation angle, $\theta_{1,2}$.
This allows us to write the planar curve's radius 
\begin{equation}\label{eqncylrad}
r = \frac{\lvert \lperp_{1,2} \rvert }{2 \sin(\theta_{1,2}/2)}.
\end{equation}
Furthermore, the dependence between the two lattice vectors shows $\lvert \lperp_2 \rvert = \lvert \lperp_1 \rvert\csc(\theta_1 / 2) \sin(\theta_2 / 2)$ so that this radius is indeed unique.
Hence our periodic origami sheets can be used to define a continuous cylinder whose symmetry axis is given by $\shat$ and whose radius of curvature is given by $\kappa = 1/r$ from Eqn.~\ref{eqncylrad}.
Since the sign of this term depends on the definition of the rotation axis, we define a more robust notion of direction via $\text{sign}(\kappa) = \text{sign}\bigg( \S_1 \l_2 \cdot (\l_1 \times \l_2) \bigg) \text{sign}\bigg( \theta_2 \bigg) = \text{sign}\bigg( \S_2 \l_1 \cdot (\l_1 \times \l_2) \bigg) \text{sign}\bigg( \theta_1 \bigg)$.

\section{Screw-periodic vertex positions}
Here, we show how to compute vertex positions in a screw-periodic lattice.
Given the position of each vertex in the unit cell, denoted by $\r_i$, we can compute the position of an arbitrary vertex by summation of all edge vectors traveled along to reach it via

\begin{equation}\label{eqnvertexpos}
\r_i(\n) = \sum_{n'=0}^{n_1-1} \S_1^{n'} \l_1 + \S_1^{n_2} \sum_{n'=0}^{n_1-1} \S_2^{n'} \l_2 + \S_1^{n_1} \S_2^{n_2} \r_i.
\end{equation}

\noindent The rotation matrices can be summed component-wise since any rotation matrix satisfies $\S^n(\theta) = \S(n \theta)$.
Without loss of generality, let us take $\shat = \hat{z}$.
We can then rewrite the terms $\cos(n \theta)$ and $\sin(n \theta)$ using Euler's formula and simply sum the exponents of complex numbers $\sum_{n'=0}^{n-1} w^{n'} = \frac{1 - w^n}{1 - w}$.
After simplification we have
\begin{align}\label{eqnrotsum}
\begin{split}
\bar{\S}(n) & \equiv \sum_{n'=0}^{n-1} \S^{n'} =
\begin{bmatrix}
\overline{\cos}(n \theta) & -\overline{\sin}(n \theta) & 0 \\
\overline{\sin}(n \theta) & \overline{\cos}(n \theta) & 0 \\
0 & 0 & n 
\end{bmatrix},\\
\overline{\cos}(n \theta) & \equiv \sum_{n'=0}^{n-1} \cos(n \theta) = \frac{1}{2} \bigg(1 - \cos(n \theta) + \cos(\frac{\theta}{2}) \sin(n \theta)\bigg), \\
\overline{\sin}(n \theta) & \equiv \sum_{n'=0}^{n-1} \sin(n \theta) = \frac{1}{2} \bigg(\cos(\frac{\theta}{2}) - \cos(\frac{\theta}{2} - n \theta) \bigg) \csc(\frac{\theta}{2}),
\end{split}
\end{align}
where we drop the subscripts for readability.

\section{Linear vertex constraint}
Here, we perform a first-order expansion of the Belcastro-Hull vertex condition, 

\begin{equation}\label{eqnbhconstraint}
\F_i = \prod_{(i,j)} \R_z(\alpha_{(i,j,k)}) \R_x(\rho_{(i,j)}) = \mathbf{I},
\end{equation}

\noindent to construct the linear vertex constraint in agreement with the main text.
Consider an infinitesimal change to the fold angles $\rho_{(i,j)} \rightarrow \rho_{(i,j)} + \phi_{(i,j)}$ where each $\lvert \phi_{(i,j)} \rvert << 1$.
The rotation matrix $\R_x(\rho_{(i,j)})$ then becomes $(\mathbf{I} + \phi_{(i,j)} \bsigma_x) \R_x(\rho_{(i,j)})$ where $\bsigma_x$ is the infinitesimal generator of rotations about the $x$-axis.
Expanding the product around vertex $i$ to first-order yields a sum of products where each term with coefficient $\phi_{(i,j)}$ shifts the location of $\bsigma_x$ so that it lies on the left of $\R_x(\rho_{(i,j)})$.
To the left of this rotation generator is the product $\bigg( \prod_{(i',j')}^{(i,j-1)} \R_z(\alpha_{(i',j',k')}) \R_x(\rho_{(i',j')}) \bigg) \R_z(\alpha_{(i,j-1,k)})$ where we use $(i,j-1)$ to denote the face-sharing edge clockwise to $(i,j)$.
As in the construction above of Eqn.~\ref{eqnbhconstraint}, this is simply the similarity transform, $\P_j$, that maps maps $\hat{x}$ to $\hat{r}_{(i,j)}$.
Moreover, since the fold angle assignment satisfies the Belcastro-Hull vertex condition the product to the right of $\bsigma_x$ is exactly the inverse of this similarity transformation.
Hence, the sum over these products of rotations become a sum over infinitesimal rotations about $\hat{r}_{(i,j)}$.
Then, using the skew-symmetry of these infinitesimal rotations we are able to rewrite the linear vertex constraint as given in the main text Eqn.~\ref{eqnlinfold}.
Note, however, we are only able to choose the direction of one edge and one face within the sheet to lie along the Cartesian basis vectors.
Hence, this condition can vary by some rotation acting uniformly on each unit vector for the remaining vertices.
This has no effect on the solutions, $\phi_{(i,j)}$, which satisfy the condition, but changes the basis in which the left nullspace (vertex zero modes) is written.

\section{Quadratic vertex constraint}
Here, we perform a second-order expansion of the Belcastro-Hull vertex condition, Eqn.~\ref{eqnbhconstraint}, to construct the quadratic vertex constraint in agreement with the main text.
Expanding $\R_x(\rho_{(i,j)})$ to second order adds an extra term $(\mathbf{I} + \phi_{(i,j)} \bsigma_x + \frac{1}{2} \phi_{(i,j)}^2 \btau_x) \R_x(\rho_{(i,j)})$ for
\begin{equation}
\btau_x \equiv
\begin{bmatrix}
0 & 0 & 0 \\
0 & -1 & 0 \\
0 & 0 & -1
\end{bmatrix}.
\end{equation}
Then expanding the Belcastro-Hull vertex condition and again using the fact that finite rotations perform a similarity transform mapping the $x$-axis onto the relevant edge vector, we obtain the sum $\sum_{(i,j)} \sum_{(i,k); k>j} \phi_{(i,j)} \phi_{(i,k)} \bsigma_{(i,j)} \bsigma_{(i,k)} + \frac{1}{2} \sum_{(i,j)} \phi_{i,j}^2 \btau_{(i,j)} = \mathbf{0}$.
This simplifies considerably using bra-ket notation to
\begin{align}\label{eqngenericquadfold}
\sum_{(i,j)} \sum_{\substack{(i,k) \\ k>j}} \phi_{(i,j)} \phi_{(i,k)} \Big( \ket{\r_{(i,j)}} \bra{\r_{(i,k)}} - \braket{\r_{(i,j)} | \r_{(i,k)}} \mathbf{I} \Big) + \frac{1}{2} \sum_{(i,j)} \phi_{(i,j)}^2 \Big( \ket{\r_{(i,j)}} \bra{\r_{(i,j)}} - \mathbf{I} \Big) = \mathbf{0},
\end{align}
using $k > j$ to signify edges $(i,k)$ always follow counter-clockwise to edge $(i,j)$.
We can then consider diagonal and off-diagonal components of this matrix constraint separately by introducing superscripts $\mu, \nu$ to denote components of the edge vectors.

Along the diagonal $\mu = \nu$ and the first sum is equivalent to
\begin{align}
\begin{split}
\sum_{(i,j)} \sum_{\substack{(i,k) \\ k>j}} \phi_{(i,j)} \phi_{(i,k)} \Big( \ket{\r^{\mu}_{(i,j)}} \bra{\r^{\mu}_{(i,k)}} - \braket{\r_{(i,j)} | \r_{(i,k)}} \Big) &
= \frac{1}{2} \Bigg( \sum_{(i,j)} \sum_{(i,k)} \phi_{(i,j)} \phi_{(i,k)} \Big( \ket{\r^{\mu}_{(i,j)}} \bra{\r^{\mu}_{(i,k)}} - \braket{\r_{(i,j)} | \r_{(i,k)}} \Big) \\ 
& - \frac{1}{2} \sum_{(i,j)} \phi_{(i,j)}^2 \Big( \ket{\r^{\mu}_{(i,j)}} \bra{\r^{\mu}_{(i,j)}} - 1 \Big) \Bigg).
\end{split}
\end{align}
The double summation of the first term can be computed independently which then necessarily vanishes when the $\phi_{(i,j)}$ satisfy the linear vertex condition

\begin{equation}\label{eqnlinfold}
\sum_{(i,j)} \phi_{(i,j)}(\n) \hat{r}_{(i,j)}(\n) = \mathbf{0}.
\end{equation}

\noindent The single summation exactly cancels the second term of Eqn.~\ref{eqngenericquadfold} implying the diagonal automatically vanishes for linear folding motions.

The off-diagonal components are not trivially satisfied, but they can be simplified into the form given the main text.
We again rewrite the double-summation
\begin{align}
\begin{split}
\sum_{(i,j)} \sum_{\substack{(i,k) \\ k>j}} \phi_{(i,j)} \phi_{(i,k)} \ket{\r^{\mu}_{(i,j)}} \bra{\r^{\nu}_{(i,k)}} 
= \sum_{(i,j)} \sum_{(i,k)} \phi_{(i,j)} \phi_{(i,k)} \ket{\r^{\mu}_{(i,j)}} \bra{\r^{\nu}_{(i,k)}} \\
- \sum_{(i,j)} \sum_{\substack{(i,k) \\ k>j}} \phi_{(i,j)} \phi_{(i,k)} \ket{\r^{\nu}_{(i,j)}} \bra{\r^{\mu}_{(i,j)}} 
- \sum_{(i,j)} \phi_{(i,j)}^2 \ket{\r^{\mu}_{(i,j)}} \bra{\r^{\nu}_{(i,j)}}.
\end{split}
\end{align}
Similarly to the diagonal components, the first double-sum vanishes for linear folding motions that satisfy Eqn.~\ref{eqnlinfold}.
Adding the remaining term in Eqn.~\ref{eqngenericquadfold} reveals this constraint is skew-symmetric.
It is hence satisfied when the difference with its transpose vanishes which is exactly the cross product definition 
\begin{equation}
\sum_{(i,j)} \sum_{\substack{(i,k) \\ k>j}} \phi_{(i,j)} \phi_{(i,k)} \hat{r}_{(i,j)} \times \hat{r}_{(i,k)} = \mathbf{0}.
\end{equation}
Finally, since $\phi_{(i,j)}$ satisfy the linear vertex condition we have $\sum_{k>j} \phi_{(i,k)} \hat{r}_{(i,k)} = - \sum_{k \leq j} \phi_{(i,k)} \hat{r}_{(i,k)}$ so this may be rewritten as given in the main text 

\begin{equation}\label{eqnquadfold}
\sum_{(i,j)} \delta \phi_{(i,j)} \hat{r}_{(i,j)} + \sum_{(i,j)} \phi_{(i,j)} \bigg( \sum_{\substack{(i,k) \\ k<j}} \phi_{(i,k)} \hat{r}_{(i,k)} \bigg) \times \hat{r}_{(i,j)} = \mathbf{0}.
\end{equation}

\noindent The interior sum gives the angular velocity $\bomega_{(i,j,j')} - \bomega_{(i,i_1,i_2)}$ so that the quadratic term gives the sum over changes in orientation of unit vectors $\rhat_{(i,j)}$ with the face $(i,i_1,i_2)$ fixed.
However, as stated in the main text, since the $\phi_{(i,j)}$ satisfy the linear vertex equation, Eqn.~\ref{eqnlinfold}, we can add the constant $\bomega_{(i,i_1,i_2)} - \bomega_{(1,2,3)}$ to Eqn.~\ref{eqnquadfold} at every vertex without changing the result.
Hence, the quadratic term is equivalent to summing over the changes in edge directions $\sum_{(i,j)} \delta \rhat_{(i,j)} \phi_{(i,j)}$.

\section{Second-order folding motions via mechanical duality}

As discussed in the main text and the previous appendix, second-order folding motions $\delta \phi_{(i,j)}$ exist provided that they satisfy the following relation to the first-order motions $\phi_{(i,j)}$ 

\begin{equation}\label{quadfold}
\sum_{(i,j)} \delta \phi_{(i,j)} \hat{r}_{(i,j)} + \sum_{(i,j)} \phi_{(i,j)} \bigg( \sum_{\substack{(i,k) \\ k<j}} \phi_{(i,k)} \hat{r}_{(i,k)} \bigg) \times \hat{r}_{(i,j)} = \mathbf{0}
\end{equation}

\noindent at every vertex $i$ in the unit cell. This first sum can, as discussed in the main text, be expressed as the vector $\Q \delta \phi$. At first appearance, it might seem that regardless of the first-order contributions a second-order correction can be chosen to satisfy the second order corrections. However, the equilibrium matrix is not invertible, so the additional term cannot have any contribution that lies in its left nullspace. Hence, a sufficient condition for the existence of the second-order correction to exist is that the double sum above not lie in the left nullspace of the equilibrium matrix, which is also the right nullspace of the compatibility matrix. Starting from a spatially periodic configuration, that consists solely of the three Euclidean translations. The resultant condition is then simply that the above quadratic condition vanish when summed over each vertex in the unit cell:

\begin{equation}
\sum_i \sum_{(i,j)} \sum_{\substack{(i,k) \\ k<j}} \phi_{(i,j)}  \phi_{(i,k)} \hat{r}_{(i,k)}  \times \hat{r}_{(i,j)} = \mathbf{0}.
\end{equation}

Now, we may show by induction that this sum over every vertex in the interior of the unit cell may be reduced to a sum over a loop drawn around the boundary of the unit cell. Suppose that this relationship already holds true for a certain loop, as it certainly does for a loop around a single vertex, when the two scenarios are the same. Suppose that we add a single adjacent vertex $i'$. In the above equation, this would increase the total sum by an amount

\begin{equation}
\sum_{(i',j)} \sum_{\substack{(i',k) \\ k<j}} \phi_{(i',j)}  \phi_{(i',k)} \hat{r}_{(i',k)}  \times \hat{r}_{(i',j)} .
\end{equation}

The loop sum, on the other hand, would instead be increased by an amount 

\begin{equation}\label{quadfoldsimp}
\sum_{(i',j)}\sum_{\substack{(i',k) \\ k<j}} \pm \phi_{(i',j)}  \phi_{(i',k)} \hat{r}_{(i',k)}  \times \hat{r}_{(i',j)}
+ \sum_{(i'',j'')} \sum_{(i',j)}\pm \phi_{(i'',j'')}  \phi_{(i',j)} \hat{r}_{(i'',j'')}  \times \hat{r}_{(i',j)}.
\end{equation}

Here, the $\pm$ is positive when a new edge has been crossed due to the addition of the new vertex and negative for an edge that has been removed from the outer loop as it winds around the new vertex. The second sum, over every edge of the new vertex and every edge that precedes it in the loop,  may be eliminated, and the first sum may be transformed into that of the previous equation, by noting that because the first-order conditions are satisfied then $\sum_{(i,j)}\phi_{(i,j)} \hat{r}_{(i,j)} = 0$ at every vertex.

In this way, the second-order conditions may be satisfied provided that the first-order conditions satisfy

\begin{align}
    \sum_{(i,j)}\sum_{(i',j')}\phi_{(i,j)}\phi_{(i',j')}\rhat_{(i,j)} \times \rhat_{(i',j')}=0,
\end{align}

\noindent where now the sum is over ordered pairs of edges encountered on a loop drawn counter-clockwise around the unit cell.

However, such a loop may also be decomposed as successive paths along the first lattice direction, the second lattice direction, the reverse of the first lattice direction and the reverse of the second lattice direction. Expressed in terms of the lattice rotations, this is simply the second-order expansion of

\begin{align}
    \S_1 \S_2 \S_1^{-1}\S_2^{-1} = \mathbf{I}.
\end{align}

Hence, we find that any set of linear folding motions may be extended to second-order if and only if it satisfies orientation compatibility to second-order. As discussed in the main text, flat vertices enlarge the left nullspace of the equilibrium matrix and generate both additional linear modes and additional constraints, therefore leading to more complicated but not higher-dimensional sets of rigid folding motions.

\section{Vertex displacements from folding motions}
Here, we evaluate the double-summation that maps from linear folding motions to vertex displacements with the face $(1,2,3)$ of cell $\n=(0,0)$ held fixed.
The angular velocity of face $(i,j,k)$ in cell $\n$ is given by the sum $\bomega_{(i,j,k)}(\n) = \sum_{(i',j')} \phi_{(i',j')} \hat{r}_{(i',j')}$ where $(i',j')$ takes on the value for each edge which is passed on a path from the initial face to face $(i,j,k)$ in cell $\n$.
Since our folding motions are uniform and the edge vectors are screw-periodic, we can expand this sum as
\begin{equation}\label{eqnangvelexp}
\bomega_{(i,j,k)}(\n) = \bar{\S}_1(n_1) \bOmega_1 + \S_1^{n_1} \bar{\S}_2(n_2) \bOmega_2 + \S_1^{n_1} \S_2^{n_2} \bomega_{(i,j,k)},
\end{equation}
using $\bomega_{(i,j,k)}$ to denote the angular velocity of face $(i,j,k)$ in the unit cell and and $\bOmega_{1,2} = \sum_{1,2} \hat{r}_{(i,j)} \phi_{(i,j)}$ denote the cell angular velocities computed by summing along the path from face $(1,2,3)$ of the $\n=(0,0)$ cell to the same face in the $\n=(1,0)$ and $\n=(0,1)$ cells respectively.
These cell angular velocities, $\bOmega_{1,2}$, give the components of the skew-symmetric matrices $\delta \S_{1,2}$ introduced in the main text.
Furthermore, Eqn.~\ref{eqnangvelexp} is independent of the order of summation since the angular velocities, $\bomega_{(i,j,k)}$, are construction from linear folding motions, $\phi_{(i,j)}$, that satisfy Eqn.~\ref{eqnlinfold}.
The displacement of vertex $k$ in cell $\n$ is $\u_k(\n) = \sum_{(i',j',k')}^{k} \big(\sum_{(i'',j'',k'')}^{(i',j',k')} \phi_{(i'',j'')}\hat{r}_{(i'',j'')} \big) \times \r_{(i',k')}$ which we now evaluate for spatially periodic and screw-periodic configurations.

First, consider a flat (not necessarily developable) configuration in which $\r_{(i,k)}(\n) = \r_{(i,k)}$ so that the periodicity is described as would be for a conventional crystal.
The angular velocity of a face in this configuration reduces to $\bomega_{(i,j,k)}(\n) = n_1 \bOmega_1 + n_2 \bOmega_2 + \bomega_{(i,j,k)}$.
The displacement of a vertex can be expanded as
\begin{align}\label{eqnvertdisp}
\begin{split}
\u_k(\n) = n_1 \sum_{(i',j',k')}^{n_1=1} \bomega_{(i',j',k')} \times \r_{(i',k')} 
+ n_2 \sum_{(i',j',k')}^{n_2=1} \bomega_{(i',j',k')} \times \r_{(i',k')} \\ + \frac{n_1 (n_1 + 1)}{2} \bOmega_1 \times \l_1
+ \frac{n_2 (n_2 + 1)}{2} \bOmega_2 \times \l_2 + n_1 n_2 \bOmega_1 \times \l_2,
\end{split}
\end{align}
where the last term is equivalent to $n_2 n_1 \bOmega_2 \times \l_1$ by since the linear folding motions satisfy the expansion of position compatibility, Eqn.~\ref{eqnposcompat}, to first-order.
The first two terms, which are linear in $n$, can be interpretted as intracell strains while the last three terms, which are quadratic in $n$, characterize intercell curvatures.
Hence, the accumulation of cell angular velocities, $\bOmega_{1,2}$, gives rise to the cylindrical structure from the flat state.
For this reason we define the following quantities 
\begin{align}
\delta \l_{1,2} &\equiv \sum_{(i',j',k')}^{n_{1,2}=1} \bomega_{(i',j',k')} \times \r_{(i',k')}, \\
\delta \kappa^{-1}_{11,22} &\equiv \bOmega_{1,2} \times \l_{1,2}, \\
\delta \kappa^{-1}_{12}  \equiv \bOmega_1 \times \l_2 &= \delta \kappa^{-1}_{21} \equiv \bOmega_2 \times \l_1,
\end{align}
for changes to the lattice vectors $\delta \l_{1,2}$ and changes to the curvatures $\delta \kappa_{ij}$.
Moreover, $\bOmega_{1,2} = \mathbf{0}$ implies the there is no curvature in the corresponding lattice direction so that it either defines the rotation axis or the sheet remains flat.
This occurs for the planar folding motions of crease patterns such as the Miura-ori and the eggbox.

More generally, we can perform the double summation for cylindrical origami.
In this case, the face angular velocities take the form given in Eqn.~\ref{eqnangvelexp} and the edge vectors take the form $\r_{(i,j)}(\n) = \S_1^{n_1} \S_2^{n_2} \r_{(i,j)}$.
This requires defining the double sum over rotation matrices $\bar{\bar{\S}}(n) = \sum_{n'=0}^{n-1} \bar{\S}(n')$.
This matrices will take the same form as Eqn.~\ref{eqnrotsum} with its components replaced by $\overline{\overline{\cos}}(n \theta)$, $\overline{\overline{\sin}}(n \theta)$, and $n(n-1)/2$.
The double-sums over the trigonometric functions require no additional evaluation when we rewrite $\cos(\frac{\theta}{2} - n \theta) = \cos(\frac{\theta}{2}) \cos(n \theta) + \sin(\frac{\theta}{2}) \sin(n \theta)$ in $\overline{\sin}(n \theta)$ since we can simply substitute our previous results to find
\begin{align}
\begin{split}
\overline{\overline{\cos}}(n \theta) & = \frac{1}{2} \bigg( n - \overline{\cos}(n \theta) + \cos(\frac{\theta}{2}) \overline{\sin}(n \theta) \bigg), \\
\overline{\overline{\sin}}(n \theta) & = \frac{1}{2} \bigg( n \cos(\frac{\theta}{2}) - \cos(\frac{\theta}{2})\overline{\cos}(n \theta)
- \sin(\frac{\theta}{2})\overline{\sin}(n \theta) \bigg) \csc(\frac{\theta}{2}).
\end{split}
\end{align}
The vertex displacements can then be written in the similar form
 \begin{align}\label{eqncylvertdisp}
\begin{split}
\u_k(\n) &= \bar{\S}_1(n_1) \sum_{(i',j',k')}^{n_1=1} \big( \bomega_{(i',j',k')} \times \r_{(i',k')} \big)
 + \S_1^{n_1} \bar{\S}_2(n_2) \sum_{(i',j',k')}^{n_2=1} \big( \bomega_{(i',j',k')} \times \r_{(i',k')} \big) \\
&+ \bar{\bar{\S}}_1(n_1) \big( \bOmega_1 \times \l_1 \big)+ \S_1^{n_1} \bar{\bar{\S}}_2(n_2) \big( \bOmega_2 \times \l_2 \big)
 + \big( \bar{\S}_1(n_1) \bOmega_1 \big) \times \big( \S_1^{n_1} \bar{\bar{\S}}_2(n_2) \l_2 \big).
\end{split}
\end{align}
Again, the first two terms have the interpretation of intracell strains and the last three terms signify changes in intercell curvatures.

We can similarly compute changes in the lattice rotation axes after a linear folding motion.
The new axis must be invariant under the new rotation 
\begin{equation}
\S' \shat' = (\mathbf{I} + \delta \S) \S (\shat + \delta \shat) = \shat',
\end{equation}
where for the moment we drop the subscripts indicating which lattice direction these rotations are in.
To first-order, this gives
\begin{equation}
(\mathbf{I} - \S) \delta \shat = \delta \S \shat.
\end{equation}
Since $\delta \shat$ must be orthogonal to $\shat$, we can project into the plane defined by $\shat$ as we did for position compatibility, Eqn.~\ref{eqnperpposcompat}
\begin{equation}
(\mathbf{I} - \Sperp) \delta \shat^{\perp}= (\delta \S \shat)^{\perp}.
\end{equation}
By the independence of summation order in 

\begin{equation}\label{eqnangvel}
\bomega_{(i,j,k)}(\n) = \sum_{(i',j')} \phi_{(i',j')}(\n') \hat{r}_{(i',j')}(\n'),
\end{equation}

\noindent the cell angular velocities obey a similar relation to Eqn.~\ref{eqnperpposcompatsol} for the lattice vectors
\begin{equation}
\bOmega_2^{\perp}  = (\mathbf{1} - \Sperp_1)^{-1}(\mathbf{1} - \Sperp_2) \bOmega_1^{\perp},
\end{equation}
so that after inversion
\begin{equation}
\delta \shat_{1,2}^{\perp} = (\mathbf{1} - \Sperp_{1,2})^{-1} (\bOmega_{1,2} \times \shat)^{\perp},
\end{equation}
we have $\delta \shat_1^{\perp} = \delta \shat_2^{\perp}$.

\section{Origami reconstruction}\label{sec:numerics}
Here, we show how to reconstruct an origami sheet from its fold angles.
Starting with edge $\hat{r}_{(1,2)} \equiv \hat{x}$, we can obtain $\hat{r}_{(1,3)} = \R_z(\alpha_{(1,2,3)}) \hat{r}_{(1,2)}$ by choosing the normal of this first face to lie in the $xy$ plane.
We may similarly obtain successive edges, however the normal vectors of successive faces are not known.
Instead, we use the fact the rotation about some axis, $\hat{z}'$, is given $\R_{z'}(\alpha) = \R_x(\rho) \R_{z}(\alpha) \R_x^{-1}(\rho)$ when $\hat{z}'$ is related to $\hat{z}$ by a rotation of $\rho$ about $\hat{x}$.
Since we know the fold angles which relate the normal vectors on successive faces in addition to the direction of the previous edge vector, we are able to use this method to find the direction for each edge leaving a vertex.
Moreover, since the edge lengths are preserved by a rigid folding motion, once we have the directions of the edges we can find the positions of the connected vertices.
By this method, we can determine the positions of all vertices in the unit cell, as well as those which are connected to a vertex in the unit cell by a crease.

This latter fact implies we obtain information about how edges rotate betweens cells which we use to determine the lattice rotations.
To be specific, given two edges in the unit cell and their counterpart in either the cell $(1,0)$ or $(0,1)$, we can construct an orthonormal basis for a plane in either cell and define the invertible matrix
\begin{equation}\label{eqnbasis}
\B = 
\begin{bmatrix}
\hat{r}_{(i,j)} & \hat{r}'_{(k,l)} & \widehat{\r_{(i,j)} \times \r'_{(k,l)}}
\end{bmatrix}^T
\end{equation}
where we have used Gram-Schmidt orthogonalization so that $\hat{r}'_{(k,l)} \cdot \hat{r}_{(i,j)} = 0$.
The matrix in an adjacent cell must be related $\B_{1,2} = \S_{1,2} \B$ so that the cell rotations can be written explicitly as $\S_{1,2} = \B_{1,2} \B^{-1}$.
The cooresponding rotation angle is then given by $\theta_{1,2} = \arccos{(Tr \S_{1,2} - 1)/2}$ and the rotation axis is given by introducing a normalization factor, $N$, to the vector 
\begin{equation}\label{eqnrotaxis}
\shat= \frac{N}{2\sin\theta}(S_{1,2}^{32}-S_{1,2}^{23},S_{1,2}^{13}-S_{1,2}^{31},S_{1,2}^{12}-S_{1,2}^{21}),
\end{equation}
 where we use superscripts to denote components of the rotation matrix.
The lattice vectors in the unit cell are more simply computed by summing along the edges which connect the first vertex to its counterpart in the neighboring cells.
Once we have the positions for all vertices in the unit cell, the lattice vectors, and the lattice rotations we can reconstruct our periodic, cylindrical sheet.
By doing this along the trajectory, we are able to show the origami sheet rigidly folding in real time.

\section{Projection into strain space}
Here, we explain the orthonormalization used for our deformation tensor.
To construct the components of the deformation tensor used to visualize the configuration space of our rigidly foldable origami sheet we introduce the first fundamental form, $\tilde{g}_{(i,j)} = \l_i \cdot \l_j$, of the unit cell.
This quantity can be used to compute distances in reference to the initially flat sheet $ds^2 = \tilde{g}_{(i,j)} \Delta n_i \Delta n_j$, where $\Delta n_i$ is the integer number of cells translated in the $i$ lattice direction.
Changes to this quantity give a description of the stretching and shearing of the lattice vectors as the sheet evolves.
For consistency, we perform a coordinate transformation so that the first fundamental form of the flat state is the identity matrix.
We write this new quantity $g_{ij} = a^{-1}_{ik} \tilde{g}_{kl} a^{-1}_{lj}$, where $a_{ij} = \hat{e}_i \cdot \l_j$ for the unit vectors $\hat{e}_i$ lying within the plane of the flat sheet.
We can then construct the deformation tensor, $\boldsymbol{\epsilon} = \mathbf{g} - \mathbf{I}$, whose three independent components, $(\epsilon_{11}, \epsilon_{22}, \epsilon_{12})$, provide a three-dimensional coordinate system to visualize the path of our origami sheet through its configuration space.

\section{Periodic and boundary modes}

In this Appendix we briefly describe the periodic and boundary modes of mechanical frames in general and the triangulated origami sheets discussed in the main text in particular, as well as their relations to quantum systems.

This article concerns itself with characterizing the isometries of triangulated origami sheets. Because a triangular frame is rigid, each of those isometries corresponds exactly to a set of vertex positions that do not stretch any of the sides of any of the triangular faces. Consequently, we can identify the linear zero modes of our triangulated origami surface by identifying the linear zero modes of a ``ball and spring'' structure with vertices located at origami vertices and inextensible bonds along origami edges.

We may then construct a linear map from vertex displacements $\mathbf{u}_1, \mathbf{u}_2$ to the extension of the bond between them as

\begin{align}
    e= (\mathbf{u}_2-\mathbf{u}_1)\cdot \hat{r}_{12},
\end{align}

\noindent where $\hat{r}_{12}$ is the unit vector pointing from the first to the second vertex. In a periodic system such as the ones we consider here, Bloch's theorem would generally indicate that the only waveforms which could be normal modes of the system (that is, modes that oscillate at some frequency, importantly including zero modes that oscillate at zero frequency) would be those modes that scaled by phase factors between neighboring cells. In fact, before periodic boundary conditions are imposed, this ``phase factor'' need not have unit magnitude and so to represent a generalized Bloch mode we define two complex numbers, $z_1, z_2$ by which a potential normal mode may scale between the two origami lattice directions. In terms of these, we may define the rigidity map $\mathbf{R}(z_1,z_2)$ from a vector $\mathbf{u}$ describing the displacements of all the sites in the unit cell to a vector $\mathbf{e}$ describing the extensions of all the bonds (origami edges) in the unit cell:

\begin{align}
    \mathbf{e} = \mathbf{R}(z_1,z_2) \mathbf{u}.
\end{align}

Because our systems are both triangulated and periodic, this rigidity map is a square matrix. This can be seen via Euler's theorem for a polyhedron, $V+F-E = \chi$. In this case, the Euler characteristic $\chi$ vanishes, because the periodic boundary conditions give our unit cell the surface topology of a torus. Furthermore, because triangles have three edges but each edge is shared between two triangles, $E = (3/2)F$. Thus, throughout the entire unit cell there are three edges for each vertex, even if not all vertices are adjacent to six edges. Since origami vertices can be displaced in three directions while edge extensions are scalars, this ensures that the rigidity map is a square matrix. Because the matrix is square but not generally Hermitian, it can be interpreted as the Hamiltonian of a dissipative quantum system.
Further, this square matrix implies zero modes exist for generalized phase factors $z_1, z_2$ if and only if its determinant vanishes:

\begin{align}
  \textrm{det}(  \mathbf{R}(z_1,z_2) ) = 0.
\end{align}

To that end, when one of the complex numbers is held fixed, locating the zero modes amounts to finding zeros of some function $f(z)$ in the complex plane. Because of the Argument Principle of complex analysis, one may calculate the number of zero modes with $|z|<1$ by calculating the change in phase of $f(z)$ as one winds around the unit circle $|z|=1$. Physically, this establishes a bulk-boundary correspondence between the number of zero modes exponentially localized to one edge and the winding of the bulk modes, the Kane Lubensky topological invariant. 

Although Newtonian dynamics is second-order in time, Kane and Lubensky compose the vertex displacements and edge extensions into a combined vector space governed by first-order dynamics. This thus assumes the form of a Schr\"{o}dinger equation, with a Hamiltonian containing both the rigidity and equilibrium matrices, lying in the BDI symmetry class.

Generically, this can lead to excesses or deficits of zero modes in such mechanical systems. However, it has been previously observed that no such excesses seem to occur in triangulated origami. We prove this via consideration of the \emph{states of self stress}.

Just as the extensions of bonds were determined in terms of the displacements of adjoining vertices above, we may also write the total force on a vertex in terms of the sum of the tensions $t_\alpha$ and the edge orientations $\hat{r}_\alpha$:

\begin{align}
    \mathbf{f} = \sum_\alpha t_\alpha \hat{r}_\alpha.
\end{align}

Consequently, if we assume unit-stiffness springs, such that tensions are extensions  we can construct a map from the bond extensions in a unit cell to the forces on the sites as

\begin{align}
    \mathbf{f} = \mathbf{R}^T(1/z_1,1/z_2)\mathbf{e}.
\end{align}

\noindent Here the linear map, known as the equilibrium map, is the transpose of the rigidity map, evaluated as the inverses of the complex numbers, or at the opposite (complex) wavevector. States of self stress are defined as the vectors in the nullspace of this map, sets of tensions (stresses) that do not support an external stress. The reason for the inversion of the complex numbers that if a mode rescales by a factor $z_1$ as one translates one cell to the right then if a site in one cell connects to a bond in the cell to the right, the bond extensions has a factor of $z$ relative to the site displacement/force, but the site displacement/force has a factor of $1/z$ relative to the bond extension.

From the rank-nullity theorem of linear algebra, we then find that the number of zero modes at a complex wavevector is equal to the number of states of self stress at the opposite wavevector. 

However, origami has a special feature not common to generic mechanical systems. The full nonlinear Belcastro Hull can be linearized into the same form as the self stress condition (see Appendix 4 and main text). At any finite wavevector, these dihedral angle changes can be integrated to obtain vertex displacements at the same wavevector. The process also works in reverse, and hence in a triangulated origami system there is exactly one zero mode at any complex wavevector for  each state of self stress at the \emph{same} wavevector.

Combining the results of the two previous paragraphs, we then find that in triangulated origami there must be equal numbers of zero modes present on opposite edges, ensuring that the system never becomes topologically polarized in the Kane Lubensky sense.

However, this does not render origami sheets topologically trivial. In fact, they possess a $\mathbb{Z}_2$ topological invariant, in contrast with the integer invariant of Kane Lubensky polarization. This invariant represents an intriguing possibility for realizing new protected modes at surfaces and interfaces. In the absence of this hidden origami duality between zero modes and states of self stress, such an interface mode would be possible only by enforcing some spatial symmetry, as well as critical coordination.

Finally, we note that because zero modes appear at opposite complex wavevectors, the determinant of the rigidity matrix must appear as a symmetrical Laurent polynomial, as discussed in the main text. This means that the rigidity matrix is real in the Brillouin Zone. Consequently, it can be positive or negative and generically the regions of the BZ in which it shifts from positive to negative are curved one-dimensional lines hosting finite-wavevector bulk zero modes. This feature is in contrast with generic critically coordinated mechanical frames, which instead host Weyl points at zero-dimensional points in a two-dimensional BZ.

These lines have also been predicted in quantum spin systems. In kagome lattices with spins of variable orientations but fixed coupling strengths, energetic contributions can be minimized when the weighted sums of spin orientations on a single kagome triangle vanish. Hence, the ground state of such a system corresponds to orientations of fixed length edges in which trios of vectors sum to zero. These in fact correspond exactly to isometries of triangulated origami sheets, as described in the references in the main text. In the latter case, the vectors summing to zero while remaining of fixed length ensures that each triangular face remains intact and undeformed. Consequently, the zero-energy magnetic excitations of the quantum system (magnons) mirror the zero-energy deformations of the classical origami.

In summary, the spatially varying zero modes of triangulated origami exist as a special class within the broader zero modes of mechanical frames, with strikingly non-generic features protected not by spatial symmetries but by the nature of the triangulated origami as both critically coordinated and an intact surface with a guaranteed duality.

\end{document}